\documentclass{IEEEcsmag}

\jvol{00}
\jnum{00}
\paper{00}
\jmonth{December}
\jname{IEEE Pervasive Computing}
\pubyear{2022}

\let\labelindent\relax

\usepackage[colorlinks,urlcolor=blue,linkcolor=blue,citecolor=blue]{hyperref}
\usepackage[inline]{enumitem}
\usepackage{booktabs} 
\usepackage{subfiles}
\usepackage{subfig}
\usepackage[nohyperlinks, printonlyused, withpage, smaller]{acronym}
\usepackage{upmath,color}
\usepackage[british]{babel}
\usepackage{hyphenat}

\expandafter\def\expandafter\UrlBreaks\expandafter{\UrlBreaks\do\/\do\*\do\-\do\~\do\'\do\"\do\-}
\begin{document}
\newacro{esm}[ESM]{Experience Sampling Method}
\newacro{iv}[IV]{Independent Variable}
\newacro{dv}[DV]{Dependent Variable}
\newacro{lsdv}[LSDV]{Least Squares Dummy Variable Model}
\newacro{svm}[SVM]{Support-Vector-Machine}
\newacro{knn}[KNN]{K-Nearest Neighbors}
\newacro{rf}[RF]{Random-Forest}
\newacro{nb}[NB]{Multinomial Naive Bayes}
\newacro{rrr}[RRR]{Relative Risk Ratio}
\newacro{mlp}[MLP]{Multi-Layer Perceptron}

\newacro{cpd}[CPD]{Change Point Detection}

\newacro{tf-idf}[TF-IDF]{Term Frequency-Inverse Document Frequency}
\newacro{tf}[TF]{Term Frequency}
\newacro{cv}[CV]{Count Vectorization}
\newacro{w2v}[W2V]{Word2Vec}
\newacro{iqr}[IQR]{interquartile range}
\newacro{hci}[HCI]{Human-Computer Interaction}

\newacroplural{esm}[ESMs]{Experience Sampling Methods}
\newacroplural{iv}[IVs]{Independent Variables}
\newacroplural{dv}[DVs]{Dependent Variables}
\newacroplural{lsdv}[LSDVs]{Least Squares Dummy Variable Models}
\newacroplural{svm}[SVMs]{Support-Vector-Machines}

\sptitle{Special Issue - Human-Centered AI}

\title{Towards Social Role-Based Interruptibility Management}

\author{Christoph Anderson}
\affil{University of Kassel, Kassel, 34121, Germany}

\author{Judith Simone Heinisch}
\affil{University of Kassel, Kassel, 34121, Germany}

\author{Shohreh Deldari}
\affil{RMIT University, Melbourne, 3000, Australia}

\author{Flora Salim}
\affil{University of New South Wales, Sydney, 2052, Australia}

\author{Sandra Ohly}
\affil{University of Kassel, Kassel, 34121, Germany}

\author{Klaus David}
\affil{University of Kassel, Kassel, 34121, Germany}

\author{Veljko Pejovi\'{c}}
\affil{University of Ljubljana, Ljubljana, 1000, Slovenia}

\markboth{Special Issue - Human-Centered AI}{Special Issue - Human-Centered AI}

\begin{abstract}
Pervasive and ubiquitous computing facilitates immediate access to information in the sense of \textit{always-on}. Information such as news, messages, or reminders can significantly enhance our daily routines but are rendered useless or disturbing when not being aligned with our intrinsic interruptibility preferences. Attention management systems use machine learning to identify short-term opportune moments, so that information delivery leads to fewer interruptions. Humans' intrinsic interruptibility preferences - established for and across social roles and life domains - would complement short-term attention and interruption management approaches. In this article, we present our comprehensive results towards social role-based attention and interruptibility management. Our approach combines on-device sensing and machine learning with theories from social science to form a personalized two-stage classification model. Finally, we discuss the challenges of the current and future AI-driven attention management systems concerning privacy, ethical issues, and future directions.
\end{abstract}

\maketitle

\chapteri{W}hen the coronavirus emerged from a local outbreak to a worldwide pandemic in March 2020, mobile phones, smartwatches, and laptops were a driving factor that enabled work-from-home regulations. While individuals were working remotely at home, they often experienced stress, frustration, and even conflicts with family members \cite{xiao2021}. The potential of pervasive and ubiquitous computing for work and private life-related conflicts results from their ability to overcome and breach peoples' preferences and boundaries. People establish preferences and boundaries to structure their work, and private-related demands \cite{Ashforth2000}. \textit{Interruptions} imposed by work-related emails, phone calls, or reminders are potential manifestations that breach individuals' preferences and boundaries -- especially in the after-hours \cite{Derks2015}. According to a recent study with $67$ participants, higher levels of exhaustion are reported, particularly when participants had valued \textit{work} over \textit{private} matters and were using their private smartphone while working \cite{Derks2021}.

Attention and interruption management systems aim to support individuals in maintaining their interruptibility preferences \cite{Anderson2018}. They try to mitigate the net effects of interruptions such as prolonged primary task completion \cite{Bailey2006}, increased rates of self-interruptions \cite{Dabbish2011}, or errors within primary task execution. However, to date, interruptibility management is focused on particular contextual descriptors associated with events and the expectation that short-term opportunities for interruptions arise when users, for example, finish particular tasks or make transitions between physical activities \cite{Okoshi2015}. Even though research in social and behavioral science promotes the idea of distinguishing interruptions for work and private domains concerning individuals' preferences \cite{Derks2015, Derks2021, Eastgate2021}, attention management is still in its infancy when it comes to supporting individuals' holistic preferences. For example, applications such as Apple's Focus mode or Android's Do-not-Disturb mode either aim to block communication entirely, do not provide support for social roles and life domains, or still need manual configuration \cite{Gross2021}. Consequently, there is a need for human-centered AI-driven approaches and systems that support individuals' interruptibility on a broader scale. Such approaches would complement approaches based on particular events, locations, or short-term opportunities.

In this article, we build upon this idea and present our research findings on social role-based interruptibility management that focuses on the respective user -- putting their social roles and preferences in the spotlight. Thus, we first discuss role theory and boundary management representing the basic principles of our approach. We then describe our mobile sensing study with $16$ participants for $5$ weeks. In this study, we captured participants' social roles, interruptibility preferences, and other contextual information in a \textit{multi-device} environment. Personalized classification models to identify individuals' interruptibility preferences and social roles are then evaluated and combined to form our vision toward social-role-based interruptibility management. Finally, we investigate the challenges of multi-device settings in detecting social roles for interruption management before discussing implications regarding privacy, ethics, and future AI-driven attention management systems using our approach.

\section{Role Theory \& Boundary Management}\label{sec:theory}
Our vision is to integrate theories on human behavior into a human-centered attention management system. From the data captured in our mobile sensing and \ac{esm} study, we extract information on social behavior and related individual interruptibility preferences. This kind of information then serves to support users in their interruptibility. Theories on social roles and boundary management provide the theoretical background to our approach. Further, these theories guide us through our study design, the questionnaires, and the extracted features we used to model individuals' interruptibility.

\subsection{Role Theory}
People establish physical, temporal, or cognitive boundaries to shape their domains \cite{Ashforth2000}. Domains are mental constructs that sort and maintain similar and associated events according to their meaning. For instance, individuals might establish the domains of \textit{work} and \textit{private} and characterize each of the domains with colleagues, friends, and demands on the domain. Within and across domains, people enact various \textit{social roles} \cite{Ashforth2000}. Social roles are defined as characteristic social behavior \cite{Biddle1986}, or \textit{expected behavior associated with a social position} where each position has its rights and obligations \cite{Merton1957}.

\subsection{Boundary Management}
Role boundaries limit the perimeters of associated social roles \cite{Ashforth2000}. Boundaries are flexible and permeable to facilitate transitions from one social role to another. Based on the findings of \cite{NippertEng1996} stating that individuals tend to segment their domains, Ashforth et al. suggest that roles can be aligned on the continuum from high segmentation to high integration \cite{Ashforth2000}. The continuum results in three different role preferences: \begin{enumerate*} \item segmentation, \item integration, and \item combination \end{enumerate*}. Whether a person is interruptible depends on their social role and preferences. While positive effects of concurrent roles on individuals' well-being have been found, the net effect of \textit{interactions and interruptions} contradicting demands is detrimental \cite{Eastgate2021}. The adverse effects on individuals due to frequent breaches of preferences in pervasive and ubiquitous computing motivated us to build a system that focuses on its users.

\begin{table*}[!tbh]
	\centering
	\caption{Overview of extracted features. Our target variables are denoted with ($+$). Information marked with ($*$) have been manually reported via questionnaires.}
	\label{table:data_types}
	\scriptsize
	\begin{tabular}{ll}
		\toprule                                                  
		\textbf{Context}                                     & \textbf{Features}                                                                                                          \\
		\midrule
									        
		\multicolumn{1}{l}{\textit{Computer (Windows \& macOS)}}   \\ 
		\cmidrule{1-2}
		Application stream                                   & number of unique applications, \ac{cv}, \\
                                                         & \ac{tf}, \ac{tf-idf} on application sequences \\
		Keyboard                                             & number of pressed keys \{chars, control keys\}, interaction \{true, false\}                                                \\
		Mouse                                                & number of pressed buttons \{left, right\}                                                                                  \\
		Interaction                                          & the total time spend with applications prior to an ESM                                                                     \\
		\cmidrule{1-2}
		\multicolumn{1}{l}{\textit{Smartphone (Android)}}    \\ 
		\cmidrule{1-2}
		Application stream                                   & number of unique applications, CV, TF, TF-IDF on application sequences                                                     \\
		Location                                             & Pluscodes \{8 and 10\}, most frequent and last pluscodes                                                    \\
		Physical activity                                    & number of unique activities - Google API                                                                                   \\
		Ringer Mode                                          & ringer mode changes, last ringer mode                                                                                      \\
		Screen state                                         & number of states \{on, off\}                                                                                               \\
		Notification                                         & number of \{notifications received, unique applications with notification\}                                                \\
		                                                     & number of notifications received from \{family, friend, work\},                                                            \\
		                                                     & number of notifications received from contacts with hierarchical relation \{same, above, below\}                           \\
    Contact                                              & hashed contact and/or group name extracted from notification titles \\
		Application genre                                    & number of different application genres                                                                                     \\
		Interaction                                          & the total time spend with applications prior to an ESM                                                                     \\
		\cmidrule{1-2}                                                  
		\multicolumn{2}{l}{\textit{Temporal}}      \\
		\cmidrule{1-2}
		Time                                                 & part of the day \{morning, noon, afternoon, evening, night\}                                                               \\
		Day of week                                          & number of day within week \{0-6\}                                                                                          \\
		Weekend                                              & \{yes, no\}                                                                                                                \\
		\toprule
		\textbf{Self reported}                        & \textbf{Description}                                                                                                       \\
		\midrule                                              
		Social relationship$^*$                                  & Information about family, friends, colleagues and contacts with no relationship                                            \\
		Task$^*$                                                 & The task the participants had progressed the most in the last $90$ minutes \protect\cite{Trippas2019}                                                 \\
		Social role$^+$      & The social role the user was in the last $15$ minutes: \{private, work, both\}                  \\
		Interruptibility$^+$ & \protect Interruptibility preferences for the last $15$ minutes: \{private-only, work-only, both, none\} \\
		\bottomrule                                               
	\end{tabular}
\end{table*}

\section{Capturing Social Roles \& Interruptibility Preferences}
There is a need of in-situ social role and interruptibility data to investigate the importance of social roles in attention and interruptibility management. To fulfill this need, we implemented two applications -- a cross platform application for Windows and macOS and a separate mobile application for Android. Both applications feature experience sampling and continuous background sensing and were used in an \textit{in-the-wild} study to capture participants' self-reported social roles, interruptibility preferences along with their locations and device usage.

\subsection{Procedure}
Experience sampling within our applications is based on two different approaches inspired by \cite{Berkel2017}. The first approach is based on a fixed schedule of $90$ minutes, asking participants about their current social role and interruptibility preferences. These questionnaires are only scheduled between $7$am and $10$pm. The second approach is event-based, showing additional questionnaires after participants interact with their phones for more than $10$ minutes. We set a minimum time of $30$ minutes between event-based questionnaires. When a questionnaire is issued, individuals can select between two social roles, namely \textit{work} and \textit{private}, as they represent the main domains differing between \textit{being at work} and \textit{private-related behavior}. \textit{Private} comprises \textit{home} and \textit{social}, including interactions with, e.g., the parents, the partner, pursuing hobbies \cite{NippertEng1996}. Contrary to this, \textit{work} represents all work-related activities, such as being at work or communicating with a colleague. As individuals may encounter difficulties when choosing between \textit{work} and \textit{private}, we added \textit{both} as the third selection in our questionnaires. When being asked about their current social role, individuals can also choose between four different interruptibility preferences: interruptible for \begin{enumerate*} \item \textit{private}, or \item \textit{work}-related interruptions, for \item \textit{both} or \item \textit{none}\end{enumerate*}. Finally, we issue questionnaires on peoples' relationships and -- if applicable -- their hierarchical relationship to a contact. These questionnaires are shown to participants as soon as they receive at least $5$ notifications from their contacts.

\begin{table*}[tbh]
	\centering
	\caption{Distribution of participants' \ac{esm} answers per device and class. In total, we received $3255$ valid questionnaires answers.}
	\label{table:answering_rates}
	\begin{tabular}{@{}llccccccccc@{}}
		\toprule
		& &\multicolumn{4}{c}{\textbf{Interruptibility}} &  & \multicolumn{3}{c}{\textbf{Social Role}}   & \\ 
    \cmidrule(lr){3-6} \cmidrule(lr){7-10} 
		\textbf{Device} & \textbf{No. ESM}  & Private & Work    & Both    & None    &    & Private & Work    & Both              \\ \midrule
		Smartphone & 2344 ($72.01\%$)    & 40.57\% & 14.46\% & 31.14\% & 13.82\%     & & 56.31\% & 28.29\% & 15.40\%   \\
		Computer    & 911 ($27.99\%$)     & 15.70\% & 36.66\% & 24.48\% & 23.16\%  & & 26.68\% & 55.32\%     & 18.00\%  \\ \bottomrule
	\end{tabular}
\end{table*}

Each application features background services that, if available, keep track of location updates, physical activities, interactions with applications, notifications, and the device's state (e.g., screen status, ringer modes). The captured data is regularly uploaded to our university server through an encrypted channel. A list of captured information and target variables is shown in Table \ref{table:data_types}.

\subsection{Dataset}
Data collection was carried out for $5$ weeks. The participants received information on using our applications, their rights (e.g., erasing their collected data on request), and privacy protection measures. Our privacy officer and ethics committee approved consent forms and data collection procedures. Overall, we captured data from $16$ participants -- $13$ male and $3$ female. At the time of our data collection, participants were between $19$ and $41$ years old (M = $31.44$ and SD = $5.17$ years). The study population comprised junior and senior academics and technical staff members. Their major tasks involved experiments, writing research papers, proposals, and the documentation of technical matters. Participants lived in five different countries on two different continents. We captured data from $14$ Android, $12$ Windows, and $2$ macOS devices. In total, $3255$ out of $10701$ questionnaires (answering rate = $30.41 \%$) were answered. The overall answering rate is comparable to other \ac{esm}-based studies within the field of interruptibility~\cite{Pejovic2014}. A distribution of \ac{esm} answers per device and class is shown in Table \ref{table:answering_rates}.

\subsection{Feature Engineering \& Classifiers}
Table \ref{table:data_types} shows the list of features we used to train and validate our machine-learning models. We extracted features within periods of $15$ minutes before an \ac{esm} questionnaire. The features encompass temporal, location, and application-based features, commonly used in attention management systems \cite{Anderson2018}. We further computed features based on self-reported information -- for example, the relationship to a contact. We also extracted application streams -- long sequences of application usage prior to an \ac{esm}. The motivation behind extracting application streams is that they might contain unique patterns related to a particular social role. To form homogeneous application streams across devices, we first grouped applications within the application stream. As some applications appear in every application stream, we removed applications containing less or no information about the device interaction. Such applications included launchers, system updates, or home screens. All remaining applications preceding a \textit{\ac{esm}} questionnaire were assigned to a unique sequence with a self-reported social role and interruptibility preference. Application sequences are textual representations. Such representations need to be transformed into sparse matrices for machine learning. Common approaches to transform textual representations into sparse matrices are \ac{tf}, \ac{tf-idf}, \ac{cv}, or neural networks such as \ac{w2v}. After a preliminary investigation, we chose \ac{cv} that extracts the number of application occurrences within sequences. In addition to application-based features, we extracted features related to keyboard and mouse events.

\subsection{Classifiers}
We chose popular classifiers from the field of attention and interruption management \cite{Anderson2018}. In particular, we evaluated \ac{mlp}, \ac{knn}, logistic regression, ridge classifier, and decision tree. To extend our evaluation, we integrated ensemble-based learning, namely, AdaBoost and \ac{rf}. We used a classifier that predicts the most frequent class as the baseline. To provide a unified representation to our classifiers, we decided to apply a binary encoding scheme to self-reported roles and interruptibility preferences. In particular, we encoded interruptibility to fall in the categories \begin{enumerate*} \item \textit{interrupt-private -- \{yes, no\}}, and \item \textit{interrupt-work -- \{no, yes\}} \end{enumerate*}. Consequently, the interruptibility of \textit{both} and \textit{none} are encoded as interrupt-private and interrupt-work as \textit{\{yes, yes\}} and \textit{\{no, no\}}, respectively. The same encoding was performed for social role answers. Note that the evaluation of our models was carried out on the original classes. We chose the \textit{weighted f1-score} to evaluate our \textit{personalized} classification models. This metric combines both precision and recall with the support of each class. Classifiers are trained on a randomized train \& test split.

\section{Interruptibility \& Social Role Classification}
In this section, we report on the results of our interruptibility and social role models. People may not use their phones during work or their work computer while being private. In such situations, we cannot collect interactions with both devices at the same time. Therefore, we trained our classifiers separately as we were interested in whether our models would be comprehensive enough to detect social roles and individuals' interruptibility under these circumstances. We first investigate the performance for classifying four different interruptibility preferences, namely \textit{work}, \textit{private}, \textit{both}, and \textit{none}. We then move on and report on our social role classification models.

\begin{figure*}[tb]
    \centering
    \subfloat[\centering \label{fig:interruption_phone} Phone-based features]{\includegraphics[width=.5\textwidth]{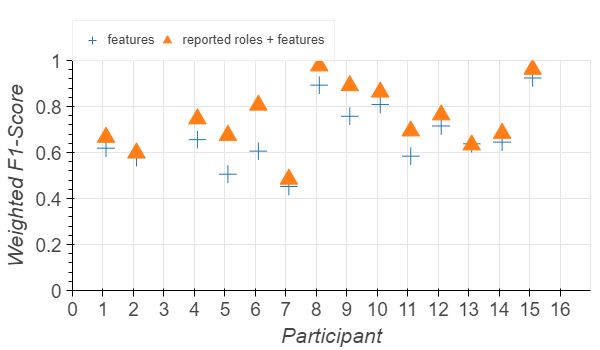}}%
    \subfloat[\centering \label{fig:interruption_desktop} Computer-based features]{\includegraphics[width=.5\textwidth]{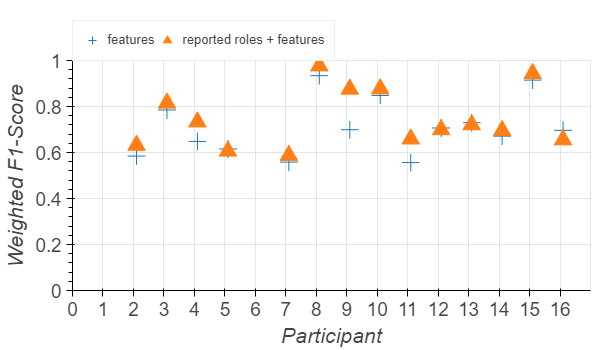}}%
    \caption{Weighted f1-scores of the Random Forest classifier for inferring individuals' interruptibility.}
    \label{fig:interruption_participants}
\end{figure*}

\subsection{Classifying individuals' interruptibility}\label{sec:interruptibility}
We built interruptibility models on common temporal and location-based features used in previous studies \cite{Okoshi2015}. Furthermore, we added the number of unique applications, physical activities, pressed keys, or the currently selected ringer mode to the feature set. Before we detail the results per device and participant, we present general classification performances computed over all participants and devices. We computed the $25th$, $50th$, and $75th$ percentiles and the mean of weighted f1-scores of all participants and classifiers. All classifiers outperformed the baseline. The median interquartile f1-score of all classifiers ranged from $0.58$ to $0.67$ (baseline $0.44$). The \ac{knn} and Adaboost classifier showed outliers below the $25th$ percentile, whereas the logistic regression and ridge classifier achieved lower minimum and higher maximum f1-scores. Overall, the \ac{rf} showed the most stable results with a median interquartile f1-score of $0.67$.

To investigate whether incorporating social roles within interruptibility models enhances the classification performance of individuals' interruptibility, we added participants' self-reported roles to our machine-learning models. The classification performance of all classifiers is improved by adding self-reported roles. The most significant improvement was observed for the ridge classifier. A conducted paired t-test confirms a statistically significant improvement of the \ac{rf} classifier. By setting the significance level $\alpha$ to $0.05$, we note that the classification of individuals' interruptibility preferences based on features only ($M = 0.70$, $SD = 0.13$) improves significantly by adding information on self-reported social roles ($M = 0.74$, $SD = 0.13$) as an additional feature ($t = -4.37, p \leq0.05$). The improvement has a medium effect size with a Cohen's d metric of $-0.36$. Figure \ref{fig:interruption_participants} shows the classification results of the \ac{rf} classifier per participant with and without self-reported social roles. We note individual differences in the improvement when incorporating self-reported social roles. As shown in Figure \ref{fig:interruption_phone} and Figure \ref{fig:interruption_desktop}, participant $15$ shows a comparable performance but no improvement. However, the classification performance for participant $5$ improves when including self-reported social roles on phone-based but not for desktop-based features. The results indicate the importance of incorporating social roles within interruptibility classification.

\subsection{Classifying individuals' social roles}
To augment future interruptibility models with individuals' enacted roles, we first need to extract their social roles through machine-learning. Therefore, we trained and tested the same classifiers as above on all features (see Table \ref{table:data_types}) to extract social roles -- \textit{private}, \textit{work}, and \textit{both}. We first present classification performances computed over all participants and devices. We then detail classification results per device. Analog to our interruptibility classification results, we computed the $25th$, $50th$, and $75th$ percentiles and the interquartile mean of obtained f1-score.

\begin{figure*}[t]
    \centering
    \includegraphics[width=\textwidth]{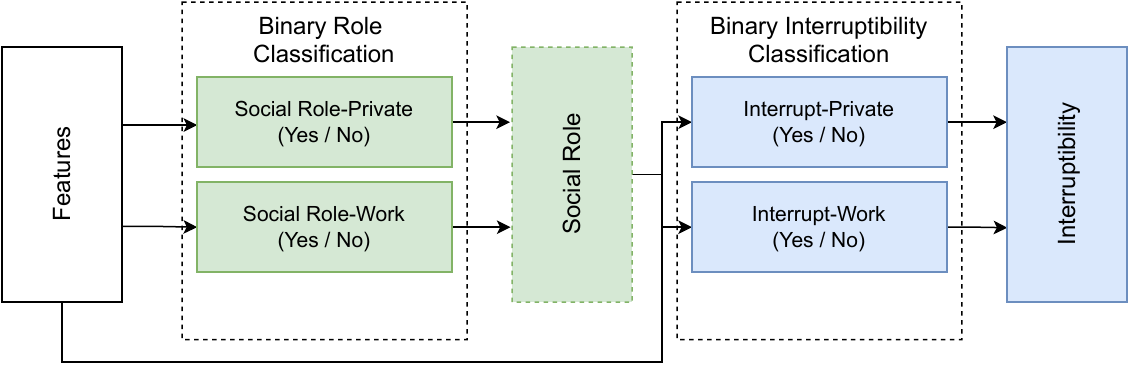}
    \caption{Two stage binary classification approach for inferring roles and interruptibility preferences.}
    \label{fig:two_stage_classifier}
\end{figure*}

We note that all classifiers perform better than the baseline. The median interquartile of the evaluated classifiers ranged from $0.76$ to $0.83$. The baseline classifier achieved a median interquartile f1-score of $0.45$. The logistic regression showed a maximum of $0.93$ and a corresponding minimum of $0.59$ weighted f1-score. Over all participants, the \ac{rf} achieves the highest median f1-score of $0.83$ for inferring social roles. The mean classification accuracy was $0.81$. Considering the performance for each class, we note that \textit{private} and \textit{work} achieved a recall of $0.83$ and $0.65$ when training the models on phone-based features only. For desktop-based features, we achieved a recall of $0.84$ and $0.71$. In both cases, the performance of the social role \textit{both} with a recall of $0.32$ and $0.25$ was underwhelming. A potential explanation is that \textit{both} represents the combined role of \textit{private} and \textit{work} and, therefore, is harder to detect.

\section{Towards Social Role-Based Interruptibility Classification}
\begin{figure*}[tbh]
    \centering
    \subfloat[\centering \label{plt:stage_results_phone} Phone-based features ]{\includegraphics[width=.5\textwidth]{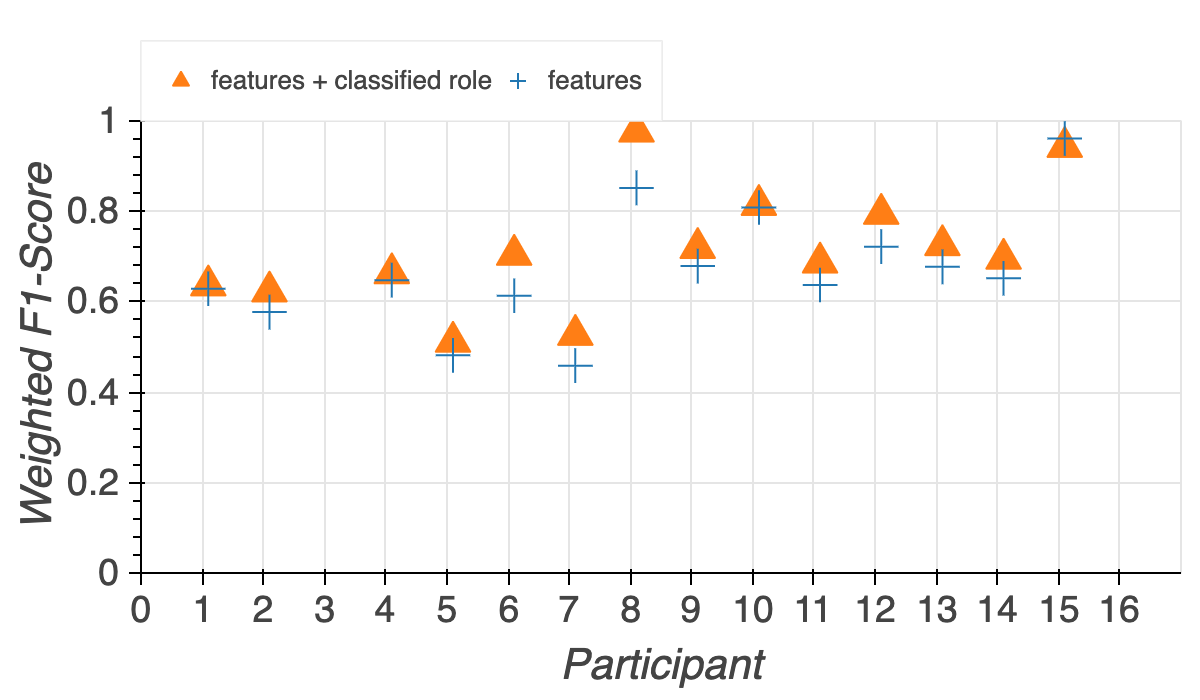}}%
    \subfloat[\centering \label{plt:stage_results_desktop} Computer-based features ]{\includegraphics[width=.5\textwidth]{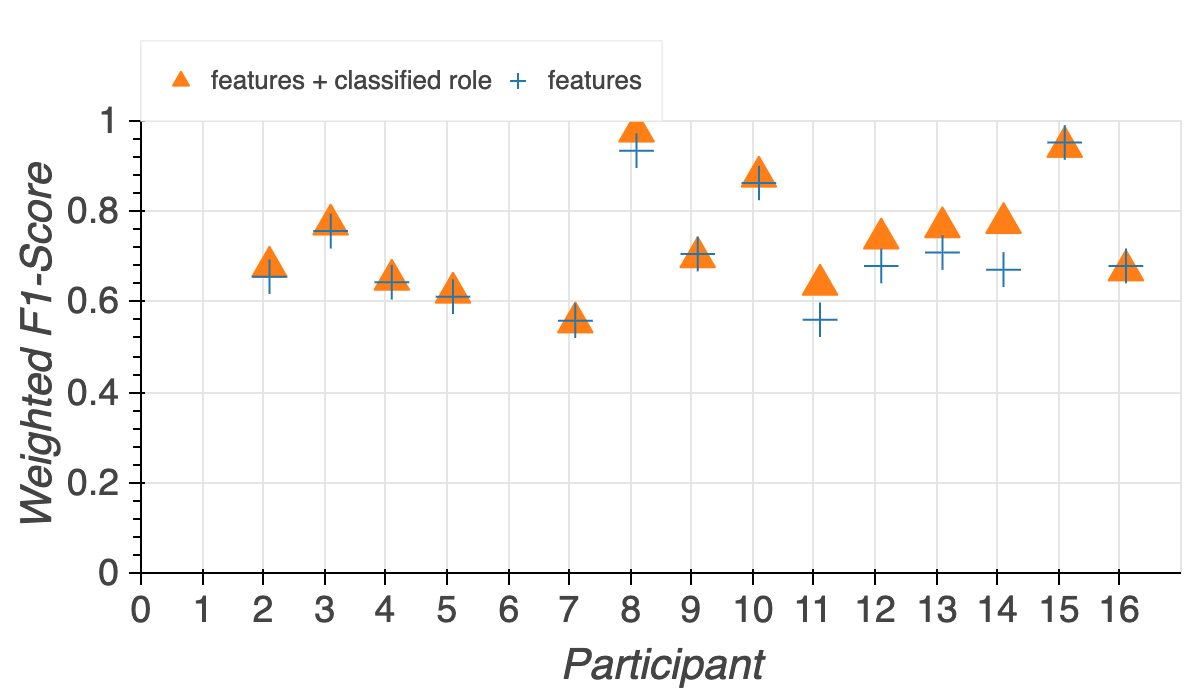}}%
    \caption{Weighted F1-scores of the \ac{rf} classifier for social role-based interruptibility classification.}
    \label{fig:result_stage}
\end{figure*}

The results from our interruptibility and social role classification models motivated us to combine both models. Therefore, we designed a social role-based interruptibility classification approach. This approach focuses on individual social behavior and related preferences to infer individuals' interruptibility. It is based on individuals' intrinsic interruptibility preferences – established for and across different social roles. The design of our social role-based interruptibility classifier is shown in Figure \ref{fig:two_stage_classifier}. Individuals' private- and work-related roles are classified in the first stage. This newly gained information is added to the available feature set from Table \ref{table:data_types} and fed to the binary interruptibility classifier in the second stage. Our two-stage model uses the \ac{rf} classifier, which showed promising results in our previous models. The interruptibility models in the second stage then perform binary classification of individuals' interruptibility preferences. The binary result is then decoded to represent our former four interruptibility preferences -- \textit{private}, \textit{work}, \textit{both}, and \textit{none}. Figure \ref{fig:result_stage} shows the results of the social role-based interruptibility classifier per participant and device. We note that the biased classified social role still improves the classification of individual interruptibility preferences. As shown in Figure \ref{plt:stage_results_phone}, all participants except participant $15$ achieve higher f1-scores for detecting their individual interruptibility preferences if using social role-based interruptibility models. When using computer-based features, all participants achieve higher or at least equal f1-scores for social role-based interruptibility models compared to traditional models, as shown in Figure \ref{plt:stage_results_desktop}.

We conducted a paired t-test to investigate if our two-stage interruptibility classification model performs differently than those interruptibility models that have been trained on temporal and application-based features. Our social role-based interruptibility models ($M = 0.73$, $SD = 0.13$) perform better than traditional interruptibility models ($M = 0.70$, $SD = 0.13$) setting $\alpha = 0.05$ ($t = -5.21, p \leq0.05$). The effect size is small to medium with a Cohen's d metric of $-0.23$. There were no statistical differences between interruptibility models evaluated on classified or self-reported social roles.

\section{Multi-Device Classification}\label{sec:multi_device}
In this section, we investigate the impact of multi-device data on social role and interruptibility classification. We conducted multiple paired t-tests and set $\alpha$ to $0.05$. Further, we use the \ac{rf} classifier, which showed promising results. As shown in Table \ref{table:t_test_phone}, we note that the combination of features from phones and computers ($M = 0.78, SD = 0.08$) significantly improves the classification for the \textit{private} role compared to using only phone-based features ($M = 0.76, SD = 0.12$). The effect is significant ($t = -2.63, p \leq0.05$). The same observation applies to the \textit{work} role where phone and computer-based features improve the classification performance ($t = -5.32, p \leq0.05$). Further, the combination of both devices only improves the interruptibility inference for \textit{private} related matters. No significant effect is observed for the interruptibility preference \textit{work}. Considering the classification of social roles and interruptibility preferences using only computer-based features (Table \ref{table:t_test_phone}), we note that the mean classification performance for being interruptible for \textit{private} related matters ($M = 0.75, SD = 0.17$) as well as for \textit{work} ($M = 0.46, SD = 0.42$) is comparable to the results of the combination of computer and phone-based features. The same observation applies to the mean classification results for the \textit{private} and \textit{work} social role. To investigate these observations further, we applied a fixed-effects multinomial logistic regression to analyze dependencies between interruptibility preferences and social roles. We modeled interruptibility preferences as \acp{dv} and social roles as \acp{iv}.

Further, temporal features from our feature set were added as time-varying covariates -- leading to $4$ different models. Significance levels and estimated coefficients remained stable -- even after integrating time-variant variables in our models. Our results indicate that participants engaged in a private role were more willing to be interrupted by private-related matters. In contrast, participants were either more interruptible for work-related matters or not interruptible while working. Furthermore, the dependencies of interruptibility preferences and roles primarily suggest a within-role interruptibility. Participants are more interruptible for events originating from the role they are currently enacting than from different roles, indicating a tendency towards \textit{segmentating} matters of their private and work-related roles.

\begin{table*}[tbh]
	\centering
	\caption{Impact of devices on the classification of social roles and interruptibility preferences.}
	\scriptsize
	\label{table:t_test_class}
	\subfloat[Phone vs. Combination]{
		\label{table:t_test_phone}
		\begin{tabular}{p{.08\textwidth}llll}
			\toprule                                                  
			\textbf{Variable} & \textbf{Phone} & \textbf{Combination} & \textbf{t-statistic} & \textbf{$\alpha$} \\
			\midrule
			\multicolumn{4}{l}{\textbf{Social Roles}}    \\ 
			private           & $M = 0.76$     & $M = 0.78$           & $-2.63$              & $\leq.05$         \\
			                  & $SD = 0.12$    & $SD = 0.08$          &                      &                   \\
			                  &                &                      &                      &                   \\
			work              & $M = 0.78$     & $M =0.83$            & $-5.32$              & $\leq.05$         \\
			                  & $SD = 0.12$    & $SD = 0.07$          &                      &                   \\
			                  &                &                      &                      &                   \\
			\midrule
			\multicolumn{4}{l}{\textbf{Interruptibility}}    \\ 
			private           & $M = 0.69$     & $M = 0.74$           & $-2.97$              & $\leq.05$         \\
			                  & $SD = 0.20$    & $SD = 0.17$          &                      &                   \\
			                  &                &                      &                      &                   \\
			work              & $M = 0.43$     & $M = 0.45$           & $-0.89$              & $0.39$            \\
			                  & $SD = 0.37$    & $SD = 0.37$          &                      &                   \\
			                  &                &                      &                      &                   \\
			\bottomrule                                               
		\end{tabular}}
	\subfloat[Computer vs. Combination]{
		\label{table:t_test_desktop}
		\begin{tabular}{p{.08\textwidth}llll}
			\toprule                                                  
			\textbf{Variable} & \textbf{Desktop} & \textbf{Combination} & \textbf{t-statistic} & \textbf{$\alpha$} \\
			\midrule
			\multicolumn{4}{l}{\textbf{Social Roles}}    \\ 
			private           & $M = 0.79$       & $M = 0.78$           & $1.67$               & $0.10$            \\
			                  & $SD = 0.08$      & $SD = 0.08$          &                      &                   \\
			                  &                  &                      &                      &                   \\
			work              & $M = 0.84$       & $M=0.83$             & $1.11$               & $0.27$            \\
			                  & $SD = 0.06$      & $SD = 0.07$          &                      &                   \\
			                  &                  &                      &                      &                   \\
			\midrule
			\multicolumn{4}{l}{\textbf{Interruptibility}}    \\ 
			private           & $M = 0.75$       & $M = 0.74$           & $0.75$               & $0.46$            \\
			                  & $SD = 0.17$      & $SD = 0.17$          &                      &                   \\
			                  &                  &                      &                      &                   \\
			work              & $M = 0.46$       & $M = 0.45$           & $0.15$               & $0.87$            \\
			                  & $SD = 0.42$      & $SD = 0.37$          &                      &                   \\
			                  &                  &                      &                      &                   \\
			\bottomrule                                               
		\end{tabular}}
\end{table*}

\section{Discussion}
Although our results indicate that information on individuals' enacted social roles can improve interruptibility classification, there are still open challenges to be considered. Our results are based on a mobile sensing study with $16$ participants over $5$ weeks. Features extracted on application usage, contacts, and the relationship to these contacts are subject to each participant. Therefore, it is challenging to generalize our results. Still, our classification results align with research findings in social and behavioral science showing that interruptions breaching individuals' preferences have adverse effects on their well-being \cite{Derks2021}, and concepts are needed to negate them \cite{Winter2010}.
Further, the target variables we used within the classification were biased among the study population. Individuals frequently selected \textit{private} or \textit{work} as a social role when asked on their smartphone or computer, respectively. Under these circumstances, our classification performance showed robust results regarding the f1-score as a combination of precision and recall. Classification models, in addition, outperformed the baseline classifier that predicted each individual's most frequent interruptibility preference. However, the distribution of our target variables might explain the multi-device t-test results in Table \ref{table:t_test_class}. As participants frequently selected \textit{work} as a social on their computers, the classification performance for work increases when combining features taken from both devices -- as computer-based features might contain more descriptive features for \textit{work}.

\section{Implications}
Our results suggest that the classification of social roles can enhance interruption management systems by covering individuals' life domains. Our approach is a complementary addition to well-established techniques based on breakpoints \cite{Okoshi2015} or mental workload \cite{Zueger2018}. In the future, social role-based attention and interruption management can be combined with short-term-based approaches in a \textit{staged system}. Such a staged system can infer individuals' enacted social roles and related interruptibility preferences while identifying short-term moments for immediate interruptions. In the first stage, the system provides information on users' receptivity toward interruptions related to their more general preferences. If users are segmenting demands, they may not be interruptible by notifications from friends or reminders from private-related applications while working. Relevant interruptions can be identified by mapping the origin of interruptions to individuals' current roles. If individuals' preferences and interruption origin coincide, short-term-based approaches further identify opportune moments for supporting individuals' fine-grained interruptibility. Interruptions might then be deferred to breakpoints or moments with a low mental workload and handled according to common strategies -- negotiated, scheduled, immediate, mediated \cite{Brixey2007}.\\Our experimental results might prove helpful in future work environments and organizations where work- and private-related interruptions are prevalent. While working, a social role-based interruption management system enforcing an individual's segmentation preferences can defer private-related interruptions to when employees start enacting their private role again. The system might also allow interruptions for an employee's private roles when applying an integration preference. Therefore, the system provides a means for organizations to employ regulations and policies to support their employees' work-life balance and well-being by reducing unwanted interruptions \cite{Derks2021}.

\subsection{Ethics \& Privacy}
There is an imminent risk of misuse and abuse of information on social roles for workplace surveillance and productivity assessments. These risks multiply as soon as detailed and fine-granular roles (e.g., \textit{parent, student, supervisor,} or \textit{colleague}) become available. With such a fine granularity, detailed profiles on social identities become practicable, including how people structure and organize demands in their environment. Consequently, the risk and chances of social role-based interruption management systems have to be weighted.

A socio-technical design approach could mitigate risks related to misuse and workplace surveillance. A key concept of socio-technical design is that the performance of a to-be-designed system depends on the joint optimization of technical \textit{and} social subsystems. Social subsystems include employees' and employers' interests and requirements that have to be weighted and equally considered in a workplace environment. The risk of misuse and workplace surveillance would then be addressed by system design. The use of multi-device data to build meaningful machine learning models bears additional privacy issues. To protect the exploitation of sensitive information, privacy-preserving machine learning models and methods related to differential privacy \cite{Dwork2006} might help. Furthermore, we propose processing the data anonymously and in compliance with GDPR directly on the user's device.

\section{Conclusion}
Pervasive and ubiquitous computing facilitates immediate access to information in the sense of \textit{always-on}, including notifications and interruptions. Attention management systems aim to address the challenge of defining short-term opportune moments for interruptions using machine learning and AI so that information delivery leads to fewer distractions. This work investigated the applicability of social role-based interruption management systems centered on the human being. Based on role theory and boundary management findings, we investigated the influence of social roles on the two-stage classification of four different interruptibility preferences.

A paired t-test confirmed that information on individuals' current social roles -- private, work, both -- significantly improves conventional interruptibility models. We combined social role and interruptibility classification in a novel two-stage classification model based on this finding. In the first stage, individuals' private- and work-related roles are classified to improve the recognition of the individual interruptibility preference (\textit{private-only}, \textit{work-only}, \textit{both}, and \textit{none}) in the second stage. Our results suggest that individuals' intrinsic interruptibility preferences -- often established for and across different social roles and life domains -- improve existing interruptibility approaches.

\section{Acknowledgements}
We thank the participants of our study. This research was partially funded by the German Research Foundation project "NORA" (44141629), by the Slovenian Research Agency project "Context-Aware On-Device Approximate Computing" (J2-3047), and by RoboTrust, a project of the Centre Responsible Digitality. Flora Salim was supported by the Australian Research Council (ARC) Discovery Project DP190101485, Humboldt Foundation, and Bayer Foundation.

\bibliographystyle{IEEEtran}
\bibliography{IEEEabrv,bibliography}

\begin{thebibliography}{10}
\providecommand{\url}[1]{#1}
\csname url@samestyle\endcsname
\providecommand{\newblock}{\relax}
\providecommand{\bibinfo}[2]{#2}
\providecommand{\BIBentrySTDinterwordspacing}{\spaceskip=0pt\relax}
\providecommand{\BIBentryALTinterwordstretchfactor}{4}
\providecommand{\BIBentryALTinterwordspacing}{\spaceskip=\fontdimen2\font plus
\BIBentryALTinterwordstretchfactor\fontdimen3\font minus
  \fontdimen4\font\relax}
\providecommand{\BIBforeignlanguage}[2]{{%
\expandafter\ifx\csname l@#1\endcsname\relax
\typeout{** WARNING: IEEEtran.bst: No hyphenation pattern has been}%
\typeout{** loaded for the language `#1'. Using the pattern for}%
\typeout{** the default language instead.}%
\else
\language=\csname l@#1\endcsname
\fi
#2}}
\providecommand{\BIBdecl}{\relax}
\BIBdecl

\bibitem{xiao2021}
Y.~Xiao, B.~Becerik-Gerber, G.~Lucas, and S.~C. Roll, ``Impacts of working from
  home during covid-19 pandemic on physical and mental well-being of office
  workstation users,'' \emph{Journal of Occupational \& Environmental
  Medicine}, vol.~63, no.~3, p. 181–190, Mar. 2021.

\bibitem{Ashforth2000}
B.~Ashforth, G.~Kreiner, and M.~Fugate, ``\BIBforeignlanguage{English (US)}{All
  in a day's work: Boundaries and micro role transitions},''
  \emph{\BIBforeignlanguage{English (US)}{Academy of Management Review}},
  vol.~25, no.~3, pp. 472--491, Jul. 2000.

\bibitem{Derks2015}
D.~Derks, D.~van Duin, M.~Tims, and A.~B. Bakker, ``Smartphone use and
  work–home interference: The moderating role of social norms and employee
  work engagement,'' \emph{Journal of Occupational and Organizational
  Psychology}, vol.~88, no.~1, pp. 155--177, 2015.

\bibitem{Derks2021}
D.~Derks, A.~B. Bakker, and M.~Gorgievski, ``Private smartphone use during
  worktime: A diary study on the unexplored costs of integrating the work and
  family domains,'' \emph{Computers in Human Behavior}, vol. 114, p. 106530,
  2021.

\bibitem{Anderson2018}
C.~Anderson, I.~H{\"u}bener, A.-K. Seipp, S.~Ohly, K.~David, and V.~Pejovic,
  ``A {S}urvey of {A}ttention {M}anagement {S}ystems in {U}biquitous
  {C}omputing {E}nvironments,'' \emph{Proceedings of the ACM on Interactive,
  Mobile, Wearable and Ubiquitous Technologies}, vol.~2, no.~2, pp. 1--27,
  2018.

\bibitem{Bailey2006}
B.~P. Bailey and J.~A. Konstan, ``On the need for attention-aware systems:
  Measuring effects of interruption on task performance, error rate, and
  affective state,'' \emph{Computers in Human Behavior}, vol.~22, no.~4, pp.
  685--708, 2006, attention aware systems.

\bibitem{Dabbish2011}
L.~Dabbish, G.~Mark, and V.~M. Gonz\'{a}lez, ``Why do i keep interrupting
  myself? environment, habit and self-interruption,'' in \emph{Proceedings of
  the SIGCHI Conference on Human Factors in Computing Systems}.\hskip 1em plus
  0.5em minus 0.4em\relax New York, NY, USA: ACM, 2011, p. 3127–3130.

\bibitem{Okoshi2015}
T.~Okoshi, J.~Ramos, H.~Nozaki, J.~Nakazawa, A.~K. Dey, and H.~Tokuda,
  ``Attelia: {Reducing} {User}'s {Cognitive} {Load} due to {Interruptive}
  {Notifications} on {Smart} {Phones},'' in \emph{2015 {IEEE} {International}
  {Conference} on {Pervasive} {Computing} and {Communications}}.\hskip 1em plus
  0.5em minus 0.4em\relax St. Louis, MO, USA: IEEE, 2015, pp. 96--104.

\bibitem{Eastgate2021}
L.~Eastgate, A.~Bialocerkowski, M.~Hood, and P.~A. Creed,
  ``\BIBforeignlanguage{en}{Applying boundary management theory to university
  students: {A} scoping review},'' \emph{\BIBforeignlanguage{en}{International
  Journal of Educational Research}}, vol. 108, p. 101793, Apr. 2021.

\bibitem{Gross2021}
T.~Gross and A.-L. Mueller, ``Notificationmanager: Personal boundary management
  on mobile devices,'' in \emph{Human-Computer Interaction -- INTERACT 2021},
  C.~Ardito, R.~Lanzilotti, A.~Malizia, H.~Petrie, A.~Piccinno, G.~Desolda, and
  K.~Inkpen, Eds.\hskip 1em plus 0.5em minus 0.4em\relax Cham: Springer
  International Publishing, 2021, pp. 243--261.

\bibitem{Biddle1986}
B.~J. Biddle, ``Recent {Developments} in {Role}-{Theory},'' \emph{Annual Review
  of Sociology}, vol.~12, no.~1, pp. 67--92, 1986.

\bibitem{Merton1957}
R.~K. Merton, ``The role-set: Problems in sociological theory,'' \emph{The
  British Journal of Sociology}, vol.~8, no.~2, pp. 106--120, 1957.

\bibitem{NippertEng1996}
C.~E. Nippert-Eng, \emph{\BIBforeignlanguage{en}{Home and {Work}: {Negotiating}
  {Boundaries} through {Everyday} {Life}}}.\hskip 1em plus 0.5em minus
  0.4em\relax Chicago, IL, USA: University of Chicago Press, Jul. 1996.

\bibitem{Trippas2019}
J.~R. Trippas, D.~Spina, F.~Scholer, A.~H. Awadallah, P.~Bailey, P.~N. Bennett,
  R.~W. White, J.~Liono, Y.~Ren, F.~D. Salim, and M.~Sanderson, ``Learning
  about work tasks to inform intelligent assistant design,'' in
  \emph{Proceedings of the 2019 Conference on Human Information Interaction and
  Retrieval}.\hskip 1em plus 0.5em minus 0.4em\relax New York, NY, USA: ACM,
  2019, p. 5–14.

\bibitem{Berkel2017}
N.~van Berkel, D.~Ferreira, and V.~Kostakos, ``The experience sampling method
  on mobile devices,'' \emph{ACM Comput. Surv.}, vol.~50, no.~6, Dec. 2017.

\bibitem{Pejovic2014}
V.~Pejovic and M.~Musolesi, ``{InterruptMe}: {Designing} {Intelligent}
  {Prompting} {Mechanisms} for {Pervasive} {Applications},'' in
  \emph{Proceedings of the 2014 {ACM} {International} {Joint} {Conference} on
  {Pervasive} and {Ubiquitous} {Computing}}.\hskip 1em plus 0.5em minus
  0.4em\relax New York, NY, USA: ACM, 2014, pp. 897--908.

\bibitem{Winter2010}
J.~Winter, D.~Cotton, J.~Gavin, and Y.~D., ``Effective e-learning?
  multi-tasking, distractions and boundary management by graduate students in
  an online environment,'' \emph{Research in Learning Technology}, vol.~18,
  no.~1, Mar. 2010.

\bibitem{Zueger2018}
M.~Züger, S.~C. Müller, A.~N. Meyer, and T.~Fritz, ``Sensing
  {Interruptibility} in the {Office}: {A} {Field} {Study} on the {Use} of
  {Biometric} and {Computer} {Interaction} {Sensors},'' in \emph{Proceedings of
  the 2018 {CHI} {Conference} on {Human} {Factors} in {Computing}
  {Systems}}.\hskip 1em plus 0.5em minus 0.4em\relax New York, NY, USA: ACM,
  2018, pp. 591:1--591:14.

\bibitem{Brixey2007}
J.~J. Brixey, D.~J. Robinson, C.~W. Johnson, T.~R. Johnson, J.~P. Turley, and
  J.~Zhang, ``A {Concept} {Analysis} of the {Phenomenon} {Interruption},''
  \emph{ANS. Advances in nursing science}, vol.~30, no.~1, pp. 26--42, Jan.
  2007.

\bibitem{Dwork2006}
C.~Dwork, ``Differential privacy,'' in \emph{Automata, Languages and
  Programming}, M.~Bugliesi, B.~Preneel, V.~Sassone, and I.~Wegener, Eds.\hskip
  1em plus 0.5em minus 0.4em\relax Berlin, Heidelberg: Springer Berlin
  Heidelberg, 2006, pp. 1--12.

\end{thebibliography}

\begin{IEEEbiography}{Christoph Anderson} is a Ph.D. student at the Department for Communication Technologies, University of Kassel, Germany. He received his M.Sc. in computer science from the University of Kassel, Germany, in 2014. His research focuses on recognizing human attention and interruptibility through interdisciplinary approaches. Contact him at \href{mailto:comtec@uni-kassel.de}{comtec@uni-kassel.de}.
\end{IEEEbiography}

\begin{IEEEbiography}{Judith Simone Heinisch} is a Ph.D. student at the Department for Communication Technology at the University of Kassel, Germany. She received her B. Sc. and M. Sc. in Computer Science from the University of Kassel in 2014 and 2017. Her research areas of interest include pervasive computing, machine learning, data mining, emotion recognition, and activity recognition. Contact her at \href{mailto:judith.heinisch@uni-kassel.de}{judith.heinisch@uni-kassel.de}.
\end{IEEEbiography}

\begin{IEEEbiography}{Shohreh Deldari} received her Bachelor's degree in Computer Engineering and her Master's degree in Computer Science from Amirkabir University (Tehran Polytechnic), Iran. Shohreh is currently doing her Ph.D. in Computer Science at RMIT University, Australia. Her research interests are machine learning, self-supervised learning, wearable sensors, and ubiquitous computing. Contact her at \href{mailto:shohreh.deldari@rmit.edu.au}{shohreh.deldari@rmit.edu.au}.
\end{IEEEbiography}

\begin{IEEEbiography}{Flora Salim} Flora Salim is a professor and the Cisco Chair of Digital Transport, School of Computer Science and Engineering, UNSW Sydney. She is a member of the Australian Research Council (ARC) College of Experts and a Chief Investigator of ARC Centre of Excellence for Automated Decision Making and Society (ADM+S). Her research is in the intersection of human-centered computing and machine learning for behavior intelligence with multimodal sensor data and AI on the edge. She was a Humboldt-Bayer Fellow, Humboldt Fellow, Victoria Fellow, and ARC APDI Fellow. She obtained her Ph.D. from Monash University in 2009. Contact her at \href{mailto:flora.salim@unsw.edu.au}{flora.salim@unsw.edu.au}.
\end{IEEEbiography}

\begin{IEEEbiography}{Sandra Ohly} is a professor in business psychology at the University of Kassel, Germany, since 2010 and director of the research center for information technology design. She received her Ph.D. from the Technical University of Braunschweig, Germany, in 2005 and completed her habilitation at the Goethe University Frankfurt, Germany, in 2010. Her research focuses on well-being, creativity, and proactive behavior at work. She is also interested in affective and motivational processes, frequently using diary methods. Contact her at \href{mailto:ohly@uni-kassel.de}{ohly@uni-kassel.de}.
\end{IEEEbiography}

\begin{IEEEbiography}{Klaus David} has been a full professor since 1998 and the Communications Technology chair since 2000 at the University of Kassel, Germany. His research interests include mobile networks, applications, and context awareness. He has published more than 200 scientific articles, including three books, and has registered more than ten patents. He was the editor-in-chief of IEEE Vehicular Technology Magazine (2015-2018) and on the IEEE Vehicular Technology Society Board of Governors (2015-2017). Contact him at \href{mailto:david@uni-kassel.de}{david@uni-kassel.de}.
\end{IEEEbiography}

\begin{IEEEbiography}{Veljko Pejovi\'{c}} is an assistant professor at the Faculty of Computer and Information Science, University of Ljubljana, Slovenia. His interests include mobile computing, HCI, and resource-efficient computing. He obtained his Ph.D. in computer science from the University of California Santa Barbara, USA. Contact him at \href{mailto:veljko.pejovic@fri.uni-lj.si}{veljko.pejovic@fri.uni-lj.si}
\end{IEEEbiography}
\end{document}